\documentclass{article}
\usepackage{spconf,amsmath,amssymb,amsfonts,graphicx}
\usepackage{booktabs}
\usepackage{array}
\usepackage{multirow}
\usepackage{xcolor}
\usepackage{cite}

\title{Kapture: Capturing Cardiac Dynamics with Koopman-Governed Learning for Efficient Radar-Based Electrocardiogram Recovery}

\name{Tong~Wu$^{1}$ \quad Jing~Peng$^{1}$ \quad Ziqi~Feng$^{2}$ \quad Yuanyuan~Zhang$^{3,*}$
\thanks{This research is supported by the Marie Curie Horizon Postdoctoral Fellowship project HealRo (Grant No.~101263991).}
\thanks{\textsuperscript{*}\textit{Corresponding author: Yuanyuan Zhang} (yz52357@uga.edu).}}

\address{$^{1}$Amazon, Seattle, Washington, United States\quad $^{2}$ Equitable, Waterloo, Canada\\
$^{3}$School of Psychology \& Neuroscience, University of Glasgow, Glasgow, United Kingdom\\
\texttt{\{tongwamz,joannapj\}@amazon.com}}

\begin{document}
\ninept
\maketitle

\begin{abstract}
Millimeter-wave (mmWave) radar enables unobtrusive, contactless electrocardiogram (ECG) reconstruction for cardiac monitoring. Time-frequency spectrograms preserve fine cardiac patterns but often require large backbones to separate ECG-relevant features from respiration, motion, multipath, and subject-dependent interference. We propose \textbf{Kapture}, a parameter-efficient Koopman-governed framework that projects radar hidden states into a low-dimensional observable space and identifies a regularized full linear evolution operator from adjacent observable states. The Koopman-predicted observables refine subsequent hidden states for ECG reconstruction. To suppress predictable interference dynamics, a temporal contrastive objective pulls neighboring states together and separates non-neighbors, while reconstruction supervision preserves task relevance. Using approximately 80 minutes of quasi-static radar--ECG recordings containing realistic noise from body movements and other sources, Kapture consistently improves reconstruction across matched backbone widths, with the largest gains under aggressive compression. The compact configuration approaches the full-width reference accuracy with $68.1\%$ fewer parameters and $91.3\%$ fewer profiler-covered floating-point operations (FLOPs), while the full-width configuration delivers the strongest overall reconstruction performance. Our code will be made publicly available after potential publication.
\end{abstract}


\begin{figure*}
\centering
\includegraphics[width=0.98\textwidth]{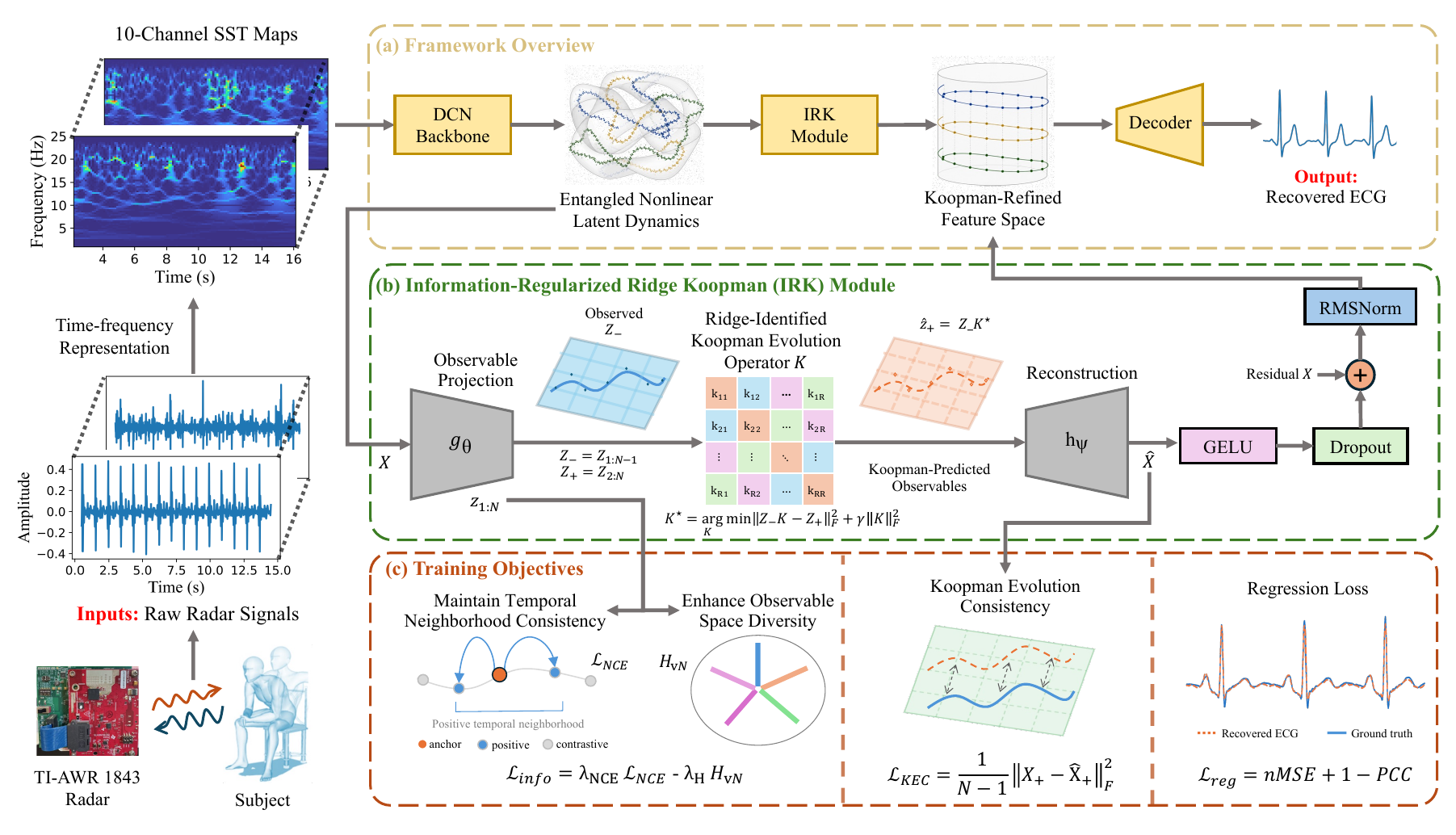}
\vspace{-8pt}
\caption{Overview of the Kapture framework: (a) Lightweight radar-to-ECG feature encoding and waveform recovery; (b) Ridge-identified Koopman feature evolution; (c) Joint objectives for Koopman consistency, information regularization, and ECG reconstruction.}
\label{fig:overview}
\end{figure*}

\section{Introduction}
\label{sec:intro}
Electrocardiogram (ECG) reflects atrial and ventricular electrical activity and is widely used for cardiac monitoring and diagnosis. Conventional acquisition relies on adhesive electrodes or wearables, which can be uncomfortable for long-term monitoring. Millimeter-wave (mmWave) radar offers a contactless alternative that senses subtle chest-wall motion, is robust to illumination and acoustic interference, and avoids camera-related privacy concerns~\cite{zhang2023overview}. Recent studies have advanced from heart-rate estimation to fine-grained ECG reconstruction~\cite{yamamoto2020ecg}. However, radar measures mechanical motion rather than electrical activity, making radar-to-ECG reconstruction a challenging cross-domain task~\cite{yamamoto2020ecg,xu2021cardiacwave}.

Early deep learning models established the feasibility of data-driven ECG reconstruction from one-dimensional radar or Doppler signals~\cite{yamamoto2020ecg,chen2025selfdenoiser,xu2021cardiacwave}. Recent studies target robustness, physiological interpretability, and limited-data learning. For example, ordinary differential equation (ODE)-based models add cardiac-cycle dynamics and morphological priors to radar-to-ECG mapping, while physics-constrained learning captures the mechanical coupling between radar-observed motion and cardiac activity to improve interpretability and robustness~\cite{zhang2024radarODE,zhang2025bloodpressure,zhang2026phybp}. Transfer learning, multi-task optimization, and data augmentation further improve reconstruction and data efficiency in small-sample settings~\cite{zhang2025rfcardi,zhang2024radarODEMTL,zhang2026phybp,zhang2025horcrux}. Together, these advances enable more robust and physiologically meaningful radar-to-ECG reconstruction under practical sensing constraints.

Despite this progress, the aforementioned approaches rely on time-frequency inputs to provide richer cardiac information than one-dimensional signals, with high-capacity backbones extracting ECG-relevant patterns from these representations~\cite{zhang2024radarODE,zhang2025rfcardi,zhang2024radarODEMTL,zhang2025horcrux,zhang2025thesis,qiu2026raecg,zhang2026conditionaldiffusion,liu2026mmjepa}. The state-of-the-art model contains about 10 million parameters~\cite{zhang2025rfcardi}, imposing substantial memory, computation, and energy demands on continuous edge deployment. Simply reducing the backbone width or latent dimension can degrade reconstruction performance because weak ECG-relevant patterns may be discarded together with respiration, multipath, and other interference sources~\cite{chu2025vessel,xiong2025mamba,visreason2026,li2025hypdae}. Prior work suggests that constraining hidden-state dynamics can address this trade-off by preserving temporally meaningful information within compact representations~\cite{brunton2021modern,chu2024vessel}.

Koopman theory provides such a constraint by lifting nonlinear dynamics into an observable space where temporal evolution can be approximated by a linear operator~\cite{brunton2021modern,xiong2026unified}. We project radar hidden states into a low-dimensional observable space and identify a Koopman operator from adjacent states through regularized ridge regression. Its predicted successor observables are mapped back to the backbone space and directly refine the feature sequence consumed by the ECG decoder. However, linear evolution alone may preserve predictable interference patterns rather than ECG-relevant cardiac dynamics. We therefore use the information noise-contrastive estimation (InfoNCE) loss~\cite{oord2018representation} to align temporal neighbors while contrasting non-neighboring states. ECG reconstruction supervision maintains task relevance, while von Neumann entropy is evaluated as an optional anti-collapse regularizer~\cite{cheng2025information}. Our main contributions are summarized as follows:
\begin{itemize}
\item We propose \textbf{Kapture}, a parameter-efficient framework governed by Koopman dynamics for radar-to-ECG reconstruction. It identifies a ridge-regularized Koopman operator in a compact observable space and uses the predicted evolution to refine the decoder's feature path.
\item We introduce a temporal contrastive learning objective based on InfoNCE. Together with ECG reconstruction supervision, it promotes locally consistent cardiac dynamics while suppressing interference-driven variation.
\item Experiments show that compact Kapture achieves reconstruction accuracy comparable to the full-width backbone with $68.1\%$ fewer parameters, demonstrating a favorable accuracy--efficiency trade-off.
\end{itemize}
\section{Methodology}
\label{sec:method}

As illustrated in Fig.~\ref{fig:overview}, Kapture follows a three-stage end-to-end path. First, the lightweight deformable convolutional network (DCN) backbone in Fig.~\ref{fig:overview}(a) extracts an ordered radar feature sequence. Next, the proposed information-regularized ridge Koopman (IRK) module in Fig.~\ref{fig:overview}(b) projects these features into an observable space, identifies their local evolution, and maps the predicted successors back to form the Koopman-refined feature sequence $X_K$. The ECG decoder then reconstructs the waveform from $X_K$, while Fig.~\ref{fig:overview}(c) summarizes the evolution-consistency, information-regularization, and regression objectives. The backbone and decoder follow~\cite{zhang2025rfcardi}, whereas the IRK module provides the proposed dynamical representation path.

\subsection{Radar Representation and Lightweight Feature Encoding}

To investigate efficient learning of cardiac-related representations, we adopt the open-source CFT-RFcardi model~\cite{zhang2025rfcardi} as the base architecture. Each multichannel radar measurement is converted into ten synchrosqueezing transform (SST) maps, denoted by $S$. As shown in Fig.~\ref{fig:overview}(a), the lightweight DCN backbone $E_\phi$ produces the ordered feature sequence $X=E_\phi(S)=[x_1,\ldots,x_N]\in\mathbb{R}^{B\times N\times D}$, where $B$, $N$, and $D$ are the batch size, sequence length, and feature width, respectively. The temporal order of $X$ provides the state sequence used by the IRK module; the original SST tensor is not directly involved in Koopman identification.

\subsection{Ridge-Identified Koopman Feature Evolution}

A lightweight backbone may lack the capacity to disentangle weak cardiac patterns from respiration~\cite{zhang2026dynamicunet}, motion, and other interference. Our intuition is to compensate for this limitation with temporal structure rather than additional width: cardiac-related features should evolve coherently across adjacent states. The IRK module therefore lifts backbone features into a compact observable space, models their local evolution with a linear operator, and maps the predicted next states back to refine the ECG feature path. Thus, the learned dynamics enrich the inference representation rather than serving only as an auxiliary training constraint.

First, as illustrated in Fig.~\ref{fig:overview}(b), a learned projection $g_\theta:\mathbb{R}^{D}\rightarrow\mathbb{R}^{R}$ lifts the backbone features into the observable sequence
\begin{equation}
Z=g_\theta(X)=[z_1,\ldots,z_N]\in\mathbb{R}^{B\times N\times R},
\label{eq:observable_projection}
\end{equation}
where the projection reduces the feature dimension from $D$ to $R$, thereby lowering the computational complexity of the subsequent Koopman operations. The adjacent state pairs are constructed as $Z_-=Z_{:,1:N-1,:}$ and $Z_+=Z_{:,2:N,:}$. After merging the batch and temporal dimensions, $Z_-$ and $Z_+$ are treated as matrices in $\mathbb{R}^{B(N-1)\times R}$.

Second, a linear Koopman evolution operator is estimated from the paired states through ridge regression:
\begin{equation}
\begin{aligned}
K^\star
&=\arg\min_K\ \|Z_-K-Z_+\|_F^2+\gamma\|K\|_F^2 \\
&=(Z_-^\top Z_-+\gamma I)^{-1}Z_-^\top Z_+,
\end{aligned}
\label{eq:ridge_koopman}
\end{equation}
where $\gamma>0$ stabilizes the solution when observable dimensions are correlated or weakly excited, and the resulting operator captures interactions among observable dimensions. Because the batch and temporal dimensions are merged before estimation, a single $K^\star$ is identified jointly from all adjacent state pairs in the current minibatch. The same closed-form solve is performed during training and inference; during training, the solution is detached from the gradient graph following the released ridge-identification formulation.

The successor observables are then predicted as $\widehat{Z}_+=Z_-K^\star$, and the reconstruction projection maps them back to the backbone feature space as $\widehat{X}_+=h_\psi(\widehat{Z}_+)$. As shown in Fig.~\ref{fig:overview}(b), $\widehat{X}_+$ passes through a Gaussian error linear unit (GELU) and dropout, is added to $X$ through a residual connection, and then undergoes root mean square layer normalization (RMSNorm) to produce $X_K$. This complete refinement path is active during both training and inference, so the identified dynamics directly modify the features consumed by the ECG decoder.

\subsection{Information-Regularized Representation Learning}

The IRK module is trained with evolution consistency and information regularization, as illustrated in Fig.~\ref{fig:overview}(c).

\subsubsection{Koopman Evolution Consistency and Temporal InfoNCE}

The $\mathcal{L}_{\mathrm{KEC}}$ term compares the reconstructed successor features with the corresponding backbone states:
\begin{equation}
\mathcal{L}_{\mathrm{KEC}}
=\frac{\|\widehat{X}_+-X_{:,2:N,:}\|_F^2}{B(N-1)D}.
\label{eq:koopman_evolution_consistency}
\end{equation}
Because $K^\star$ is detached, this loss trains $g_\theta$ and $h_\psi$ to construct an observable space that supports useful next-state prediction.

In addition, neighboring radar states are expected to share locally continuous cardiac information. Let $\mathcal{V}$ denote the set of valid temporal anchors. For each $t\in\mathcal{V}$, the set $\mathcal{P}_t=\{j:1\leq |j-t|\leq k\}$ contains the temporal positives, while the remaining non-anchor states form the contrastive denominator. With cosine similarity $s(\cdot,\cdot)$ and temperature $\tau$, the temporal InfoNCE loss is
\begin{equation}
\mathcal{L}_{\mathrm{NCE}}=
-\frac{1}{|\mathcal{V}|}\sum_{t\in\mathcal{V}}
\log\frac{\sum_{j\in\mathcal{P}_t}\exp(s(z_t,z_j)/\tau)}
{\sum_{j\neq t}\exp(s(z_t,z_j)/\tau)}.
\label{eq:info_nce}
\end{equation}
This objective encourages local consistency without forcing the complete sequence to share the same representation.

\subsubsection{Observable Entropy and Joint Optimization}

To measure representation diversity, let $\overline{Z}$ be the centered observable matrix after flattening the batch and temporal dimensions. The normalized covariance and its von Neumann entropy are
\begin{equation}
\Sigma=\frac{\overline{Z}^\top\overline{Z}}{BN-1},
\qquad
\rho=\frac{\Sigma}{\operatorname{tr}(\Sigma)},
\qquad
\mathcal{H}_{\mathrm{vN}}=-\operatorname{tr}(\rho\log\rho).
\label{eq:vn_entropy}
\end{equation}
Maximizing $\mathcal{H}_{\mathrm{vN}}$ prevents the observable covariance from concentrating on a single dominant direction. The training-only auxiliary objective combines the three terms as
\begin{equation}
\mathcal{L}_{\mathrm{aux}}
=\mathcal{L}_{\mathrm{KEC}}
+\lambda_{\mathrm{NCE}}\mathcal{L}_{\mathrm{NCE}}
-\lambda_{\mathrm{H}}\mathcal{H}_{\mathrm{vN}}.
\label{eq:auxiliary_loss}
\end{equation}

The ECG decoder produces $\widehat{y}=D_\omega(X_K)$ from the Koopman-refined sequence. For the min--max normalized reference ECG $y_n$, normalized mean squared error (nMSE) and Pearson correlation coefficient (PCC) define the training objectives:
\begin{equation}
\begin{aligned}
\mathcal{L}_{\mathrm{reg}}
&=\operatorname{nMSE}(\widehat{y},y_n)
+1-\operatorname{PCC}(\widehat{y},y_n),\\
\mathcal{L}
&=\mathcal{L}_{\mathrm{reg}}+\mathcal{L}_{\mathrm{aux}}.
\end{aligned}
\label{eq:joint_loss}
\end{equation}
The regression loss maintains ECG relevance, $\mathcal{L}_{\mathrm{KEC}}$ supervises feature evolution, and the information terms shape the observable sequence.

\section{Experimental Results and Evaluations}
\label{sec:exp}

\begin{table}[t]
\centering
\caption{Performance of different methods and matched Koopman ablations across backbone widths.}
\label{tab:accuracy_efficiency}
\vspace{2pt}
\setlength{\tabcolsep}{3.8pt}
\setlength{\belowbottomsep}{2pt}
\renewcommand{\arraystretch}{0.98}
\begin{tabular}{l c c c c}
\toprule
& \multicolumn{2}{c}{ECG recovery} & \multicolumn{2}{c}{Deployment efficiency} \\
\cmidrule(lr){2-3}\cmidrule(lr){4-5}
\multicolumn{1}{c}{Model} & nMSE $\downarrow$ & PCC $\uparrow$ &
Parameters $\downarrow$ & FLOPs $\downarrow$ \\
& ($\times10^{-2}$) & (\%) & ($10^6$) & ($10^9$) \\
\midrule


RaECG~\cite{qiu2026raecg}
& $1.55$ & $83.14$
& 2.887 & 0.691 \\

Con-Diff~\cite{zhang2026conditionaldiffusion}
& $1.95$ & $80.49$
& $\mathbf{0.907}$ & 11.196 \\

mmJEPA~\cite{liu2026mmjepa}
& $2.26$ & $75.24$
& $\underline{2.141}$ & 22.985 \\

\midrule

Baseline ($D{=}64$)
& $2.32$ & $79.34$
& \multirow{2}{*}{2.612} & \multirow{2}{*}{$\mathbf{0.104}$} \\

Kapture ($D{=}64$)
& $1.82$ & $82.77$
& & \\

\cmidrule(lr){1-3}

Baseline ($D{=}128$)
& $1.79$ & $82.85$
& \multirow{2}{*}{2.932} & \multirow{2}{*}{$\underline{0.218}$} \\

Kapture ($D{=}128$)
& $1.54$ & $85.10$
& & \\

\cmidrule(lr){1-3}

Baseline ($D{=}256$)
& $1.61$ & $83.47$
& \multirow{2}{*}{4.206} & \multirow{2}{*}{0.761} \\

Kapture ($D{=}256$)
& $1.51$ & $85.52$
& & \\

\cmidrule(lr){1-3}

CFT-RFcardi$^*$~\cite{zhang2025rfcardi}
& $\underline{1.50}$ & $\underline{85.55}$
& \multirow{2}{*}{9.186} & \multirow{2}{*}{2.499} \\

Kapture ($D{=}512$)
& $\mathbf{1.49}$ & $\mathbf{85.92}$
& & \\
\bottomrule
\multicolumn{5}{l}{$^{*}$ CFT-RFcardi is also the baseline model with $D{=}512$.} \\
\multicolumn{5}{l}{\parbox{0.96\columnwidth}} \\
\end{tabular}
\end{table}

\begin{table}[t]
\centering
\caption{Ablation of the loss terms for Kapture ($D=128$).}
\label{tab:loss_ablation}
\vspace{2pt}
\setlength{\tabcolsep}{3.25pt}
\setlength{\belowbottomsep}{2pt}
\renewcommand{\arraystretch}{0.98}
\begin{tabular}{l c c c c c}
\toprule
& \multicolumn{3}{c}{Loss} & \multicolumn{2}{c}{ECG recovery} \\
\cmidrule(lr){2-4}\cmidrule(lr){5-6}
Configuration & $\mathcal{L}_{\mathrm{KEC}}$ & $\mathcal{L}_{\mathrm{NCE}}$ &
$\mathcal{H}_{\mathrm{vN}}$ & nMSE $\downarrow$ & PCC $\uparrow$ \\
& & & & ($\times10^{-2}$) & (\%) \\
\midrule
Koopman only
& $\checkmark$ & -- & -- & $1.61$ & $\underline{81.47}$ \\

$+\mathcal{L}_{\mathrm{NCE}}$
& $\checkmark$ & $\checkmark$ & -- & $\underline{1.56}$ & $79.43$ \\

$+$ entropy
& $\checkmark$ & -- & $\checkmark$ & $1.60$ & $77.96$ \\

Full objective
& $\checkmark$ & $\checkmark$ & $\checkmark$ & $\mathbf{1.54}$ & $\mathbf{85.10}$ \\
\bottomrule
\end{tabular}
\end{table}

\begin{figure*}
\centering
\includegraphics[width=0.97\textwidth]{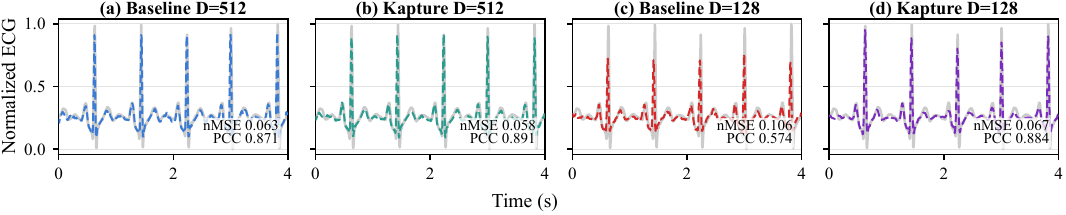}
\vspace{-8pt}
\caption{Recovered ECG with ground truth in gray: (a) - (d) Results from different models.}
\label{fig:ecg_reconstruction_result}
\end{figure*}

\begin{figure}
\centering
\includegraphics[width=0.88\columnwidth]{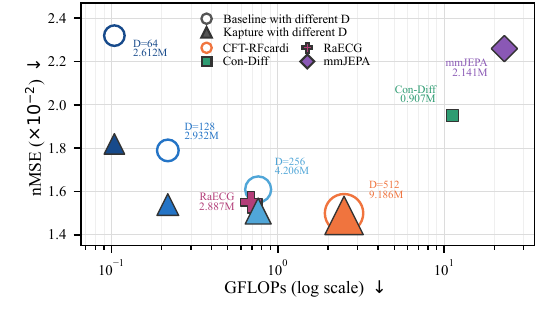}
\vspace{-8pt}
\caption{Accuracy--efficiency comparison; marker shape identifies the model variant and bubble area indicates parameter count.}
\label{fig:efficiency_bubble}
\end{figure}

\subsection{Setup and Comparisons}

We use $\gamma=0.1$, $k=3$, $\tau=0.1$, $\lambda_{\mathrm{NCE}}=0.1$, and $\lambda_{\mathrm{H}}=0.1$ throughout all experiments and set $R=16$. The evaluation uses the 81-trial, ten-subject CFT-RFcardi dataset and the original 200-Hz synchronized radar--ECG protocol~\cite{zhang2025rfcardi}. Results follow 10-fold leave-one-subject-out evaluation and report nMSE and PCC, where lower nMSE and higher PCC indicate better reconstruction. Each experiment is repeated five times, and the mean result is reported. Since all reported improvements are statistically significant, confidence intervals are omitted for brevity. We adopt CFT-RFcardi as the primary baseline because Kapture retains its backbone and ECG decoder. At a matched backbone width, this baseline is equivalent to Kapture without the proposed Koopman module, enabling a direct assessment of the module's contribution. RaECG~\cite{qiu2026raecg}, Con-Diff~\cite{zhang2026conditionaldiffusion}, and mmJEPA~\cite{liu2026mmjepa} are reproduced through the common SST--ECG interface.

\subsection{Effect of Koopman Refinement under Backbone Compression}

Table~\ref{tab:accuracy_efficiency} shows that Kapture improves both metrics at every width while the added IRK module incurs only a small overhead relative to the backbone. At $D{=}64$, it reduces nMSE by $21.6\%$ and improves PCC by 3.43 percentage points; at $D{=}128$, the corresponding gains are $14.0\%$ and 2.25 points. The gain narrows with width, supporting our premise that explicit temporal evolution is most useful when backbone capacity is constrained.

Compared with full-width CFT-RFcardi, $D{=}128$ Kapture changes nMSE only from 1.50 to 1.54 and PCC from $85.55\%$ to $85.10\%$, with $68.1\%$ fewer parameters. It also matches RaECG's nMSE, improves its PCC with roughly one-third of its profiler-covered floating-point operations (FLOPs), and avoids the substantially greater computation of Con-Diff and mmJEPA. The $D{=}512$ model gives the best overall scores but only a marginal gain over the full-width baseline, confirming that Kapture's principal advantage lies under backbone compression.

\subsection{Objective Ablation}

Table~\ref{tab:loss_ablation} isolates the auxiliary loss terms at $D{=}128$. Adding either $\mathcal{L}_{\mathrm{NCE}}$ or observable entropy alone fails to improve both reconstruction metrics: each lowers PCC relative to $\mathcal{L}_{\mathrm{KEC}}$ alone, indicating that temporal alignment or representation diversity in isolation can distort task-relevant morphology. In contrast, combining the two losses produces the lowest nMSE and highest PCC. This result suggests that the objectives are complementary: $\mathcal{L}_{\mathrm{NCE}}$ promotes locally coherent dynamics, while entropy prevents these dynamics from collapsing into an overly restricted representation.

\subsection{Qualitative Reconstruction Analysis}

Figure~\ref{fig:ecg_reconstruction_result} shows the waveform-level effect. At $D{=}512$, Kapture makes a modest correction to peak shape and amplitude, reducing nMSE from 0.063 to 0.058 and increasing PCC from 0.871 to 0.891. At $D{=}128$, the baseline retains beat timing but attenuates the dominant peaks and oversmooths local morphology, obtaining an nMSE of 0.106 and PCC of 0.574. Koopman refinement restores sharper peaks, more faithful amplitudes, and more consistent inter-beat structure, reducing nMSE to 0.067 and increasing PCC to 0.884. The compact reconstruction therefore approaches the large-backbone result in both waveform morphology and aggregate accuracy.

\subsection{Accuracy--Efficiency Trade-off}

Figure~\ref{fig:efficiency_bubble} makes the trade-off visible through the paired markers at each width. Kapture lies below its matched baseline at nearly identical horizontal positions because the added IRK module incurs only a small computational overhead relative to the backbone. The separation is largest on the low-compute side, where explicit temporal evolution is most valuable.

Increasing the width from $D{=}64$ to $D{=}128$ yields a clear accuracy gain with a small computational footprint. Beyond this point, cost rises rapidly while nMSE improves progressively less. The $D{=}128$ model is therefore a practical knee point: it remains close to full-width CFT-RFcardi while reducing parameters by $68.1\%$ and profiler-covered FLOPs by $91.3\%$. The $D{=}64$ and $D{=}512$ variants define the lowest-cost and accuracy-oriented endpoints, respectively.

\section{Conclusion}
\label{sec:conclusion}

This paper presented Kapture, a parameter-efficient radar-to-ECG framework that combines ridge-identified Koopman refinement with information-regularized representation learning. By modeling compact latent dynamics and feeding the predicted evolution back into reconstruction, Kapture improves waveform recovery, particularly under aggressive feature compression, with qualitative improvements in cardiac morphology and inter-beat consistency. The compact $D{=}128$ model approaches the full-width backbone with $68.1\%$ fewer parameters, demonstrating a favorable accuracy--efficiency trade-off. These results highlight the potential of Koopman-guided latent dynamics for resource-constrained contactless sensing. Future work will investigate robustness, cross-condition generalization, and real-time edge deployment.

\clearpage
\bibliographystyle{IEEEbib}
\bibliography{references}

\end{document}